\documentclass[aps,prl,twocolumn,10pt,showpacs,superscriptaddress,amsmath,amssymb]{revtex4}

\usepackage{graphicx}
\usepackage{dcolumn}
\usepackage{bm}
\usepackage{amsmath}
\usepackage{ulem}

\begin{document}

\title{Breakdown of scale invariance in a quasi-two-dimensional Bose gas due to the presence of the third dimension}

\author{Karina Merloti}
\author{Romain Dubessy}
\author{Laurent Longchambon}
\affiliation{Laboratoire de physique des lasers, CNRS, Universit\'{e} Paris 13, 
Sorbonne Paris Cit\'{e}, 99 avenue J.-B. Cl\'{e}ment, F-93430 Villetaneuse, France}
\author{Maxim Olshanii}
\affiliation{Department of Physics, University of Massachusetts Boston, Boston, Massachusetts 02125, USA}
\affiliation{Laboratoire de physique des lasers, CNRS, Universit\'{e} Paris 13, 
Sorbonne Paris Cit\'{e}, 99 avenue J.-B. Cl\'{e}ment, F-93430 Villetaneuse, France}
\author{H\'{e}l\`{e}ne Perrin} \email{helene.perrin@univ-paris13.fr}
\affiliation{Laboratoire de physique des lasers, CNRS, Universit\'{e} Paris 13, 
Sorbonne Paris Cit\'{e}, 99 avenue J.-B. Cl\'{e}ment, F-93430 Villetaneuse, France}

\date{\today}


%
\begin{abstract}
In this Rapid Communication, we describe how the presence of the third 
dimension may break the scale invariance in a two-dimensional Bose 
gas in a pancake-shaped trap. From  the two-dimensional perspective, the possibility of a weak spilling of the atomic density beyond the ground-state 
of the confinement alters the two-dimensional chemical potential; in turn, this correction no longer supports scale invariance. 
We compare experimental data with numerical and analytic perturbative results and find a good agreement. 
\end{abstract}

\maketitle

{\it Introduction}.
Scale invariance---the absence of a length scale associated with interactions between particles in a many-body system---has attracted 
substantial attention in the past decade. The paradigmatic example of a scale-invariant system is a three-dimensional unitary gas  
\cite{ohara2002_2179,gelm2003_011401,bourdel2003_020402,gupta2003_1723,regal2013_230404}, which constitutes a quantum interacting 
system with no governing parameters at all. At the mean-field level, the two-dimensional finite-strength-interacting 
gases  are governed by a single dimensionless coupling constant
and thus also show scale-invariance, both in the bosonic \cite{hung2011_236} and in the spin-$1/2$-fermionic  \cite{vogt2012_070404,taylor2012_135301} 
cases. 
However, here, scale invariance becomes weakly broken under quantization \cite{olshanii2010_095302}, exhibiting a quantum anomaly. 

The quantum anomaly \cite{olshanii2010_095302}---a violation of scale invariance at the microscopic level, for a \textit{genuinely two-dimensional gas}---is not the only potential cause for lifting the scale invariance in the two-dimensional (2D) case. Since, empirically, 
the 2D gas is embedded in a 3D space, the \textit{third dimension} may manifest itself as well. 
Already at the mean-field level, one can see that  
the interatomic pressure should push the atomic density beyond the ground-state of the confinement, thus modifying 
the dependence of the chemical potential on the 2D density \cite{xia2013_63}. Our Rapid Communication is devoted to the study of the consequences 
of this effect. Incidentally, the effects associated with the 2D-3D crossover are currently prominent in the physics 
of 2D Fermi gases \cite{fisher2013_023612,sommer2012_045302,zhanf2012_235302}.

The third dimension effectively introduces a new energy scale corresponding to the confinement energy quantum, $\hbar \omega_{z}$,
and a new length scale, the confinement length $\tilde{a}_{z} \equiv \sqrt{2\hbar/(m\omega_{z})}$. 
Recall that the length scale associated 
with the quantum anomaly \cite{olshanii2010_095302} is represented by the two-dimensional scattering length 
$a_{2D} = C_{2D} \tilde{a}_{z} \exp[-(\sqrt{\pi}/2)(\tilde{a}_{z}/a_{3D})]$.
Here, $a_{3D}$ is the three-dimensional scattering length and $C_{2D} = 1.47\ldots$ \cite{petrov2001}.  

In Ref.\ \cite{pitaevskii1997}, Pitaevskii and Rosch 
have shown that one of the dynamical consequences of scale invariance for a two-dimensional 
harmonically trapped short-range-interacting gas is that the frequency of the lowest monopole excitation 
acquires a universal value $\Omega_{B} = 2 \omega_r$, where $\omega_r$ is the frequency of the two-dimensional trap.  
This property has been clearly confirmed experimentally for the pancake-shaped traps \cite{hung2011_236,vogt2012_070404}, and for 
the related case of a highly elongated trap \cite{chevy2001_250402}.
In this Rapid Communication, we are investigating 
how the $\Omega_{B} = 2 \omega_r $ feature becomes violated, for a chemical 
potential becoming comparable to the vibrational quantum in the confined dimension.

{\it Perturbative corrections to the collective frequencies due to corrections to the 
equation of state}.
To compute the shift of the two-dimensional breathing frequency, we follow the general procedure 
established by Pitaevskii and Stringari \cite{pitaevskii1998_4541} for correcting the zero-temperature collective excitation 
frequencies when a small correction to the equation of state is added. 
Consider a zero-temperature gas governed by an equation of state 
$\mu(n)$, where $\mu$ is  
the chemical potential, and $n$ is the density. Consider further
small time-periodic excitation $n({\bm r},\,t) \approx \bar{n}({\bm r}) + \delta n({\bm r}) \sin(\Omega t)$  
above a steady-state density distribution $\bar{n}({\bm r})$. 
Then, the equation for finding frequencies $\Omega $ and mode functions 
$\delta n({\bm r})$ reads 
\begin{align}
& - \frac{1}{\mu'(\bar{n})} {\bm \nabla}\!\cdot\! \left(\mu'(\bar{n})^2 \bar{n} {\bm \nabla} \delta n  \right) 
\label{Sturm-Liouville}
\\
&\qquad\qquad - \frac{1}{2}  \left[ {\bm \nabla}\!\cdot\! \left( \mu''(\bar{n}) {\bm \nabla} \bar{n}^2 \right)  \right] \delta n
= m \Omega^2 \, \delta n \, ,
\nonumber
\\
&{\bm \nabla} \left[ \mu(\bar{n}({\bm r})) + V({\bm r})\right] = {\bm 0} 
\label{steady_state_density}
\,\,,
\end{align}
where $m$ is the particle mass, 
$V({\bm r})$ is an external trapping potential, $\mu'=\frac{d\mu}{dn}$, and $\mu''=\frac{d^2 \mu}{dn^2}$.

Equation (\ref{Sturm-Liouville}) is written in a manifestly Sturm-Liouville form. The induced 
inner product is 
\begin{align}
\langle \delta n_{\mbox{\scriptsize I}} | \delta n_{\mbox{\scriptsize II}} \rangle \equiv \eta \int \! d^d {\bm r} \, \mu'(\bar{n}({\bm r})) 
\delta n_{\mbox{\scriptsize I}}({\bm r}) \delta n_{\mbox{\scriptsize II}}({\bm r}) 
\,\,,
\label{inner_product}
\end{align}
where the measure $\mu'(n)$ is assumed to be everywhere positive and finite, and $\eta$ is an arbitrary positive real constant. 
Here, $d$ is the dimensionality of space. For Bose condensates, where $\mu(n) = g n$, the most 
natural choice for $\eta$ is $\eta = 1/g$: in this case, the product (\ref{inner_product}) becomes the conventional inner product 
between two functions. For a proper choice of boundary conditions at the edge of the atomic cloud,
the Liouvillean 
$
\hat{L} \equiv 
- \frac{1}{\mu'(\bar{n}) m} {\bm \nabla} \mu'(\bar{n})^2 \bar{n} {\bm \nabla}
- \frac{1}{2 m}  \left[ {\bm \nabla}\!\cdot\! \left( \mu''(\bar{n}) {\bm \nabla} \bar{n}^2 \right) \right]
$ on the left-hand side 
of Eq.\ \ref{Sturm-Liouville} becomes self-adjoint with respect to the inner product (\ref{inner_product}).
However, the boundary conditions are dictated by the physics of a particular problem at hand, and they may or may 
not support this property.

Consider now the particular case of a two-dimensional gas in a radially symmetric potential well. Assume that its chemical 
potential $\mu(n)$ is an analytic function of the density $n$, and its Taylor expansion at small density starts from 
a nonzero linear term: $\mu(n) = g n  + {\cal{O}}(n^2)$. In this case, one can show that any solution of the eigenmode 
equation (\ref{Sturm-Liouville}) behaves, at the edge of the cloud $r \approx R$, as $\delta n(r,\, \Theta) = A(\Theta) \ln(R-r) + B(\Theta) + {\cal O}(R-r) $,
where $R$ is the radius of the cloud, $(r,\,\Theta)$ are the polar coordinates, and $A(\Theta)$ and $B(\Theta)$ are arbitrary functions 
of the polar angle $\Theta$. 
The logarithmic divergence {\it per se} does not {\it a priori} mean that a particular mode is not physical: in 
particular, no divergence in the number of atoms follow. However, one can further look at the small excitations of the gas using the Lagrange representation 
of the hydrodynamic equations \cite{landau_hydrodynamics}. There, one considers the Lagrange trajectories of the particles of the constituent gas, 
${\bm r}(t,\,\tilde{\bm r})$, as functions of time $t$ and the initial coordinate $\tilde{\bm r}$. It is easy to show 
that for small amplitude excitations, the following relationship between the Euler (conventional) and Lagrange variables applies:
\begin{align}
\delta n({\bm r}) \approx {\bm \nabla}\cdot\left[ \bar{n}({\bm r}) \delta \tilde{\bm r}(\bm r)\right]
\,,
\label{Euler-Lagrange}
\end{align}
where the density is again $n({\bm r},\,t) \approx \bar{n}(r) + \delta n({\bm r}) \sin(\Omega t)$, 
and the inverse of a Lagrange trajectory, $\tilde{\bm r}(t,\, {\bm r})$, is   
$\tilde{\bm r}(t,\, {\bm r}) \approx {\bm r} + \delta\tilde{\bm r}(\bm r) \sin(\Omega t)$.
According to the relationship (\ref{Euler-Lagrange}), a logarithmic divergence of the density at the edge of the 
cloud, together with the linear decay of the steady-state density, $\bar{n}(r ) \propto R-r$, leads, at best, to a logarithmic 
divergence in $\tilde{\bm r}(\bm r)$. In turn, such a divergence would mean that the particles at the edge of the cloud 
were initially positioned infinitely far from it. To ensure the physical significance of the solutions of Eq.\ (\ref{Sturm-Liouville}), 
one would need to impose the following boundary condition on the mode functions: 
\begin{align}
&
\delta n(r,\, \Theta) = A(\Theta) \ln(R-r) + B(\Theta) + O(R-r)
\nonumber
\\
&
A(\Theta) = 0
\label{bc}
\,.
\end{align}
Furthermore, it is easy to show that the Liouvillean $\hat{L}$ is self-adjoint on the space of small excitations 
obeying the boundary condition (\ref{bc}), for the bilinear form (\ref{inner_product}) regarded as the inner product.    

The self-adjoint property of the operator $\hat{L}$ leads to {\it dramatic} simplifications of the expressions for the corrections to 
excitation frequencies under small modifications of the equation of state. Superficially, such calculation would require the knowledge of
all unperturbed mode functions and all unperturbed frequencies. But the existence of an inner product (\ref{inner_product}) with respect to 
which the unperturbed Liouvillean is self-adjoint implies---similar to what happens in quantum mechanics---that the dominant 
correction to $\Omega^2 $ depends {\it only} on the mode function of the mode whose frequency is corrected:
\begin{align}
\Delta (\Omega^2) = \langle \delta n_{0}^{}\, | \, \Delta \hat{L}\, | \,\delta n_{0}^{} \rangle 
\,\,.
\label{Delta_Omega2}
\end{align}
Here, $\delta n_{0}^{}$ is an eigenmode of the unperturbed 
Liouvillean $\hat{L}_{0}$: i.e.\ $\hat{L}_{0}\, \delta n_{0}^{} = \Omega_{0}^2 \, \delta n_{0}^{}$; the full Liouvillean
is supposed to consist of $\hat{L}_{0}$ and a small perturbation $\varepsilon \Delta \hat{L}$: 
{\it i.e.}\  $\hat{L}  = \hat{L}_{0} + \varepsilon \Delta \hat{L}$, where $\varepsilon = 1$ is a dummy ``small'' parameter;
the square frequency is assumed to be Taylor-expanded onto a power series in the powers of $\varepsilon$: 
$\Omega^2 = \Omega_{0}^2 +\varepsilon\, \Delta (\Omega^2) + {\cal O}(\varepsilon^2)$;
the mode function $\delta n_{0}$ is normalized as 
$\langle \delta n_{0}^{}\, | \, \delta n_{0}^{} \rangle = 1$. The inner product in (\ref{Delta_Omega2}) is understood as being based on the 
unperturbed measure. Notice that for the validity of the relationship (\ref{Delta_Omega2}) (unlike for the validity of the perturbation theory corrections of the higher order) it is sufficient that the unperturbed Liouvillean $\hat{L}_{0}$ alone be self-adjoint.

Next, we consider how a small perturbation $\varepsilon \, \Delta \mu(n)$ to the unperturbed equation of state 
$\mu_{0}(n)$ affects the Liouvillean $\hat{L}$.
The procedure to obtain the first-order correction $\varepsilon \, \Delta \hat{L}$  is as follows. 
There are two distinct 
contributions. One, obviously, comes from the correction to the dependence of the chemical potential on the density. 
However, the steady-state density distribution $\bar{n} = \bar{n}_{0} + \varepsilon \Delta\bar{n} + O(\varepsilon^2)$ 
also acquires an $\varepsilon^{1}$ correction
\begin{align}
\Delta\bar{n} = 
- \Delta \mu(\bar{n}_{0}) + (\Delta \mu)_{N})/ \mu_{0}'\left(\bar{n}_{0}\right)
\,\,,
\label{correction_to_density}
\end{align}
where the correction to the chemical potential
$
(\Delta \mu)_{N}
\equiv 
\left( \int \! d^d {\bm r} \, \Delta \mu(\bar{n}_{0})/\mu_{0}'(\bar{n}_{0})\right) 
/
\left( \int \! d^d {\bm r}/\mu_{0}'(\bar{n}_{0})\right)
$ is introduced to ensure that the {\it number of atoms in the perturbed system is the same as in the unperturbed one}. 
The density correction (\ref{correction_to_density}), 
in turn, independently contributes to $\hat{L}$.  
The final expression for the correction
$\Delta \hat{L}$ reads, in a compact form,
\begin{align*}
\Delta \hat{L} = 
\frac{\partial}{\partial \varepsilon}  
\hat{L}
\Big|_{\mbox{\scriptsize $
\begin{array}{l}
\mu(n) \to \mu_{0}(n) + \varepsilon \, \Delta \mu(n) + \ldots
\\
\bar{n} \to \bar{n}_{0} + \varepsilon \, \Delta \bar{n} + \ldots
\\
\varepsilon \to 0
\end{array}
$}}
\,\,.
\end{align*}

Alternatively, corrections associated with the potential changes in the number of particles can
be accounted for as follows. (i) The corresponding correction to the chemical 
potential is set to zero: 
\begin{align}
\Delta\bar{n} = 
- \Delta \mu(\bar{n}_{0}) / \mu_{0}'(\bar{n}_{0})
\,\,.
\label{correction_to_density_bis}
\end{align}
In this case the density correction (\ref{correction_to_density})
will correspond to {\it the same position of the edge of the cloud for both unperturbed and perturbed
equations of state}, but different numbers of particles. (ii) The correction to the frequency is represented by 
two terms: 
\begin{align}
\Delta (\Omega^2) = \langle \delta n_{0}^{}\, | \, \Delta \hat{L}\, | \,\delta n_{0}^{} \rangle 
-
\frac{\partial}{\partial N}(\Omega_{0}^2) \, \Delta N
\,\,,
\label{Delta_Omega2_bis}
\end{align}
with $\Delta N = -\int \! d^d {\bm r} \, \Delta \mu(\bar{n}_{0})/\mu_{0}'(\bar{n}_{0})$ being the 
correction to the number of atoms. The purpose 
of the second term in the expression (\ref{Delta_Omega2_bis}) is to ``undo'' the correction to the unperturbed frequency associated with a modification 
of the number of particles induced by the expression (\ref{correction_to_density_bis}). Note 
that this term is identically zero for Bose condensates.

{\it The confined-dimension-induced shift for a quasi-two-dimensional Bose gas}.
Let us now turn to the concrete example of breathing frequencies of a quasi-two-dimensional Bose gas in a radially 
symmetric harmonic trap. The unperturbed equation of state 
reads $\mu_{0}(n) = g n$, where $g = 4\sqrt{\pi}(\hbar^2/m)(a_{3D}/\tilde{a}_{z})$ is the two-dimensional coupling constant. 
This expression assumes that, in the confined dimension, atoms occupy the 
ground vibrational state only. A better approximation for the chemical potential 
can be obtained by using the exact ground-state chemical 
potential of an infinite two-dimensional transversally harmonic slab: 
$\mu(n) = \mu_{1D} \Big|_{g_{1D} = g_{3D}n}  - \frac{1}{2}\hbar\omega_{z}$,
where $\mu_{1D}$ is the ground-state chemical potential of the one-dimensional nonlinear Schr\"{o}dinger equation
$\left[-(\hbar^2/2m) \partial^2/\partial z^2 + m\omega_{z}^2 z^2/2 + g_{1D} |\phi(z)|^2 \right] \phi(z) = \mu_{1D} \phi(z)$,
and $g_{3D} = 4\pi\hbar^2 a_{3D}/m$.
The first two orders of the expansion of $\mu(n)$ in the powers of $g_{1D}$ read 
$
\frac{\mu(n)}{\hbar\omega_{z}}
= \frac{g n}{\hbar\omega_{z}} - \left(\frac{\sqrt{2} g_{3D}}{\tilde{a}_{z}\hbar \omega_{z}}\right)^2 |c_{2}| n^2 + \ldots 
$,
where
\begin{align}
c_{2} 
&= - \frac{3}{\pi} \sum_{m=1}^{\infty} \frac{1}{2^{m} \cdot m \cdot m!} \left[ \int_{-\infty}^{+\infty}\! d\xi e^{-2\xi^2} H_{m}(\xi)  \right]^2
\label{c2}
\\
&= -0.033\ldots
\,\,,
\nonumber
\end{align}
and $H_{m}(\xi)$ are the Hermite polynomials.
This number is also confirmed by a straightforward numerical simulation of the ground-state of the one-dimensional 
nonlinear Schr\"{o}dinger equation. 

Now, we invoke the actual unperturbed steady-state density,
$\bar{n}_{0}(r ) = \bar{n}_{0}(r\!=\!0 )(1-(r/R_{\mbox{\scriptsize TF}})^2)$ and the unperturbed 
mode function, $\delta n_{0}(r ) = \sqrt{3/\pi}\, (1- 2(r/R_{\mbox{\scriptsize TF}})^2)/R_{\mbox{\scriptsize TF}}^2$. 
Here, $R_{\mbox{\scriptsize TF}} = ((4 g N)/(\pi \omega_r^2 m))^{1/4}$ is the Thomas-Fermi radius that defines the position of the edge of the cloud,
and $N$ is the number of atoms.
Finally, applying formula (\ref{Delta_Omega2}), we get the following expression for the third-dimension-induced relative 
shift of the breathing frequency of a two-dimensional Bose gas: 
\begin{align}
\Delta \Omega_{B}/(\Omega_{B})_{0} = - \pi |c_{2}| \alpha + {\cal O}(\alpha^2) 
\,\,,
\label{Omega_B}
\end{align}
where $c_{2}$ is given by (\ref{c2}), $(\Omega_{B})_{0} = 2 \omega_r$ is the unperturbed frequency, 
and the governing small parameter $\alpha \equiv \mu(n(r=0))/(2\hbar\omega_{z})$ 
is given by the ratio between the chemical potential in the center of the trap and the minimal relative-motion energy of a two-body 
collision needed to excite any non-trivial transverse vibrations; the factor of $2$ is due to parity conservation in the two-body collisions.

\begin{figure} 
\includegraphics[scale=.6]{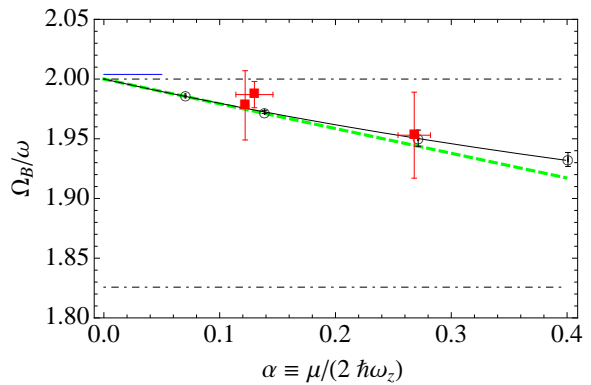}
\caption
{
(Color online)
Frequency of the breathing excitation of a two-dimensional gas, as a function of the chemical 
potential. The filled squares (red online) with error bars correspond to the experimental data. The leftmost point at $\alpha=0.122$ is an average over 
two experimental realizations: one that uses a resonant excitation scheme, and another that uses a quench (see the body of the paper). The error bars are the statistical error. The open circles 
(black online) with error bars
represent an {\it ab initio} zero-temperature three-dimensional Gross-Pitaevskii calculation; the error bars are the numerical error. The connecting line is a one-parametric exponential 
fit that crosses the $\Omega_{B} = 2\,\omega_r$ line at $\alpha=0$ and 
reaches an $\Omega_{B} = \sqrt{10/3}\, \omega_r$ plateau at $\alpha\to\infty$. The thick dashed line (green online) is given by the perturbative expression (\ref{Omega_B}) that describes 
the effect of a weak ``spilling'' of the condensate wave function along the 
confined dimension. The upper black horizontal dot-dashed line is the frequency $2\,\omega_r$ predicted by scale invariance. The lower one, $\sqrt{10/3}\, \omega_r$, is the limit of a large chemical potential, $\mu \gg \hbar\omega_{z}$, where the Thomas-Fermi profile in the confined direction is reached. The short thin solid horizontal line (blue online) reflects the effect of the quantum anomaly \cite{olshanii2010_095302}---a genuinely two-dimensional effect originating from the microscopic contributions to the macroscopic equation of state. Note that the two-dimensional chemical potential
$\mu(n)$ will differ from the full chemical potential $\mu_{3D} = \mu(n) +\frac{1}{2} \hbar\omega_{z}$ by the confinement ground-state energy.
\label{fig:_R__OmegaB_vs_alpha_2}
}
\end{figure}
%

{\it Experiment}.
We now present the experimental measurement of $\Delta \Omega_{B}$. 
$^{87}$Rb atoms are confined to the quasi-two-dimensional regime in a very anisotropic  adiabatic potential resulting from the combination of a quadrupole magnetic field and a radiofrequency (rf) field~\cite{merloti2013_033007}. The strong vertical oscillation frequency ranges from $\omega_z/(2\pi) = 1.2$ to 2.5~kHz, which corresponds to a dimensionless interaction constant $\tilde g = mg/\hbar^2$ between 0.087 and 0.12. The two-dimensional harmonic isotropic potential in the horizontal plane is very smooth: it possesses the frequency $\omega_r/(2\pi)$ ranging from 18.3~Hz to 25.0~Hz. The in-plane trap anisotropy is kept below $3\%$, which corresponds to a monopole frequency shift below $0.1\%$, well below the resolution of our measurement. This is achieved by adjusting the polarization of the rf field in order to balance the horizontal frequencies, measured at the $0.3\%$ level by recording the center-of-mass oscillations in the horizontal plane~\cite{merloti2013_033007}. The value of $\alpha$ is controlled by changing the initial atom number $N$ and the trap parameters $\omega_r$ and $\omega_z$. For each experimental point of Fig.~\ref{fig:_R__OmegaB_vs_alpha_2}, $\alpha$ is estimated from the measured atom number and trap frequencies, the chemical potential being calculated from its expression for a two-dimensional gas in the Thomas Fermi regime: $\alpha\simeq\sqrt{N\tilde g/(4\pi)}\, \omega_r/\omega_z$.

To measure $\Delta \Omega_{B} $, we proceed as follows.
(i) 
A quantum-degenerate sample of about $2\times10^4$ atoms, with a temperature of the order or below  the transverse energy $\hbar \omega_z$, is prepared following the procedure described in~\cite{merloti2013_033007}. The monopole mode is excited either by a resonant parametric modulation of the radial trapping frequency---via modulating the magnetic quadrupole field gradient---or by a sudden change in this frequency, achieved in turn through a change in the rf frequency. In both cases, the excitation process is slow as compared to the vertical trapping frequency. While in the former case the monopole mode is selected resonantly, in the latter case the symmetry of the excitation allows one to favor this mode over other modes of similar frequency, owing to the fact that their angular 
momentum is different from zero.  
(ii)  
The subsequent cloud dynamics is recorded after a given holding time in the trap. We measure the cloud size by absorption imaging along the horizontal direction after a 25~ms time-of-flight expansion. The cloud  becomes very elongated, reversing the anisotropy~\cite{merloti2013_033007}; the horizontal Thomas-Fermi radius is deduced from a bimodal fit. The monopole frequency is obtained from the time evolution of the horizontal radius.

The fitted radial cloud size exhibits time-dependent oscillations whose frequency is estimated using a sinusoidal fit, allowing one to measure the oscillation frequency of the mode. An example of such fitted oscillatory data can be found in our previous  publication, see Fig.~6 of Ref.~\cite{merloti2013_033007}. The independent measurement of $\omega_r$ gives access to the ratio $\Omega_B/\omega_r$ plotted on Fig.~\ref{fig:_R__OmegaB_vs_alpha_2}. We repeat this procedure for different values of $\alpha$.

We observe a good agreement between the experimental data and the prediction (\ref{Omega_B}). The agreement is further improved 
by employing a three-dimensional time-dependent Gross-Pitaevskii equation. Here, we use the imaginary time propagation to produce 
the steady-state, and then excite the condensate via a rapid change in the trapping frequency. A split-operator method is used at both stages.

{\it Summary and discussion}.
In this Rapid Communication, we obtain an analytic expression for an empirically relevant scale-invariance-breaking 
correction to the frequency of a two-dimensional Bose gas in a pancake-shaped trap and compare it with both the experimental data and 
numerical simulations; 
the correction is due to a weak deviation of the state of the condensate along the third, confined, dimension from the
ground vibrational state. The relative correction to the scale-invariance-dictated monopole frequency $(\Omega_{B})_{0} = 2 \omega_r $
is of the order of the ratio between the two-dimensional chemical potential $\mu$ 
and the energy of transverse confinement: 
$\Delta \Omega_{B}/(\Omega_{B})_{0} \sim \mu/\hbar\omega_{z} \sim n a_{3D} \tilde{a}_{z}$ (see Eq.\ \ref{Omega_B}). 

For typical experimental conditions, the three-dimensional effect 
we study in this Rapid Communication dominates over the purely two-dimensional {\it genuine}, quantum anomaly
\cite{olshanii2010_095302}. However, observation of the later is not out of reach.
The requirement that the frequency shift caused by the quantum anomaly,
$(\Delta \Omega_{B})_{\mbox{\scriptsize anomaly}}/(\Omega_{B})_{0} = (1/(4 \sqrt{\pi}))\,(a_{3D}/\tilde{a}_{z})$, exceeds the 
shift (\ref{Omega_B}) leads to the following upper bound on the number 
of atoms: 
\begin{align}
&
N < 
\frac{1}{16\, (c_{2})^2 \,\pi^{5/2}}
\,
\left(\frac{\omega_{z}}{ \omega_r}\right)^2
\,
\frac{a_{3D}}{\tilde{a}_{z}}
\label{visibility_of_quantum_anomaly}
\\
&
\qquad\qquad\qquad\qquad\qquad\qquad
  \simeq
3.3
\,
\left(\frac{\omega_{z}}{\omega_r}\right)^2
\,
\frac{a_{3D}}{\tilde{a}_{z}}
\,.
\nonumber
\end{align}
Consider as an example an ensemble of rubidium 87 atoms ($a_{3D} = 5.3$~nm) in a pancake trap with the transverse and in-plane 
frequencies $\omega_z = 2\pi \times 12$~kHz and $ \omega_r = 2\pi \times 25$~Hz respectively. For this set of parameters, the genuine quantum anomaly 
becomes detectable for atoms numbers belonging to the following window:
$$
4 \ll N < 3\times 10^4
\,.
$$
While the upper bound is the numerical value of the bound in 
(\ref{visibility_of_quantum_anomaly}), 
the lower bound stems from the additional requirement that
the Thomas-Fermi approximation be applicable: $\tilde{g} N \gg 1$. For instance, a group of $N=10^4$ atoms would constitute 
a Thomas-Fermi cloud whose breathing frequency will experience a quantum anomaly shift that will exceed the third-dimension-induced 
shift (\ref{Omega_B}) by approximately a factor of $\sqrt{3}$.
   

\begin{acknowledgments}

We are grateful to Eric Cornell, Vanja Dunjko, and Vincent Lorent for enlightening discussions on the subject, and to Aur\'elien Perrin for assistance on the absorption imaging. 
Laboratoire de physique des lasers is UMR 7538 of CNRS and Paris 13 University. 
We acknowledge support from the Institut Francilien de Recherche sur les Atomes Froids (IFRAF).
M.O. was supported by grants from the Office of Naval Research
(No. N00014-12-1-0400) and the National Science Foundation (No. PHY-1019197).

\end{acknowledgments}



\end{document}